\documentclass[10pt]{amsart}
\usepackage{amsmath,amssymb,amsthm,amscd,mathrsfs}
\usepackage{latexsym,amsmath,amssymb,graphicx}
\usepackage[all]{xy}
\usepackage{color}
\usepackage{epsf}
\xyoption{v2} \xyoption{2cell} \xyoption{dvips}
\CompileMatrices \numberwithin{equation}{section}
\newtheorem{prop}{Proposition}[section]

\newtheorem{exam}[prop]{Example}

\numberwithin{equation}{section}
\newcommand{\be}{\begin{equation}}
\newcommand{\ee}{\end{equation}}

\newcommand{\IP}{\mathbb{P}}%{{\relax{\rm I\kern-.18em P}}}

\newcommand\IZ{\mathbb {Z}}

\newcommand\IQ{\mathbb {Q}}
\newcommand{\IC}{\mathbb{C}}

\newcommand{\IR}{\mathbb{R}}

\newcommand{\ba}{\begin{array}}
\newcommand{\ea}{\end{array}}

\newcommand{\CV}{{\mathcal V}}

\newcommand{\CB}{{\mathcal B}}

\newcommand{\bal}{\begin{aligned}}
\newcommand{\eal}{\end{aligned}}

\newcommand{\CZ}{{\mathcal Z}}

\newcommand{\longto}{\longrightarrow}

\newcommand{\ch}{{\mathrm{ch}}}

\newcommand{\CO}{{\mathcal O}}

\newcommand{\CE}{{\mathcal E}}

\newcommand{\CH}{{\mathcal H}}

\newcommand{\CM}{{\mathcal M}}

\newcommand{\CC}{{\mathcal C}}

\newcommand{\calP}{{\mathcal P}}

\CompileMatrices \LaTeXdiagrams \UseAllTwocells
\newdimen\tableauside\tableauside=1.0ex
\newdimen\tableaurule\tableaurule=0.4pt
\newdimen\tableaustep
\def\phantomhrule#1{\hbox{\vbox to0pt{\hrule height\tableaurule width#1\vss}}}
\def\phantomvrule#1{\vbox{\hbox to0pt{\vrule width\tableaurule height#1\hss}}}
\def\sqr{\vbox{%
  \phantomhrule\tableaustep
  \hbox{\phantomvrule\tableaustep\kern\tableaustep\phantomvrule\tableaustep}%
  \hbox{\vbox{\phantomhrule\tableauside}\kern-\tableaurule}}}
\def\squares#1{\hbox{\count0=#1\noindent\loop\sqr
  \advance\count0 by-1 \ifnum\count0>0\repeat}}
\def\tableau#1{\vcenter{\offinterlineskip
  \tableaustep=\tableauside\advance\tableaustep by-\tableaurule
  \kern\normallineskip\hbox
    {\kern\normallineskip\vbox
      {\gettableau#1 0 }%
     \kern\normallineskip\kern\tableaurule}%
  \kern\normallineskip\kern\tableaurule}}
\def\gettableau#1 {\ifnum#1=0\let\next=\null\else
  \squares{#1}\let\next=\gettableau\fi\next}
\title{BPS states and the $P=W$ conjecture}
\author[W.-y. Chuang, D.-E. Diaconescu, 
 G. Pan,]{W.-y. Chuang${}^{1}$, 
D.-E. Diaconescu${}^2$, 
G. Pan${}^2$}
\address{${}^1$ Department of Mathematics, National Taiwan University, Taipei, Taiwan}
\address{${}^2$ NHETC, Rutgers University\\
Piscataway, NJ 08854-0849 USA}

\begin{document}
\begin{abstract}
A string theoretic framework is presented for the work of 
Hausel and Rodriguez-Vilegas as well as de Cataldo, Hausel and Migliorini on the cohomology of character varieties. 
The central element of this construction is  an identification of the cohomology of the Hitchin moduli 
space with BPS states in a local Calabi-Yau threefold. This is a summary of several talks given 
during the Moduli Space Program 2011 at Isaac Newton Institute.

\end{abstract} 

\maketitle

\section{Introduction}\label{one}
Consider an M-theory compactification on a smooth projective 
Calabi-Yau threefold  $Y$. M2-branes wrapping holomorphic 
curves in $Y$ yield supersymmetric BPS states in the five dimensional 
effective action. These particles are electrically charged under the 
low energy $U(1)$ gauge fields. The lattice of electric charges is 
naturally identified with 
second homology lattice $H_2(Y,\IZ)$.
Quantum states of massive particles in five dimensions 
also form multiplets of the little group $SU(2)_L\times SU(2)_R\subset 
Spin(4,1)$, which is the stabilizer of the time direction in $\IR^5$. 
The unitary irreducible 
representations of $SU(2)_L\times SU(2)_R$
may be labelled by pairs of 
half-integers $(j_L,j_R)\in \big({1\over 2}\IZ\big)^2$, which are the left, respectively right moving spin quantum numbers.
In conclusion, the space of five dimensional BPS states 
admits a direct sum decomposition
\[
\CH_{BPS}(Y)\simeq \bigoplus_{\beta\in H_2(Y,\IZ)} 
\bigoplus_{j_L,j_R\in {1\over 2}\IZ} \CH_{BPS}(Y,\beta,j_L,j_R).
\]
The refined  Gopakumar-Vafa invariants are the BPS degeneracies 
\[
N(Y,\beta,j_L,j_R) ={\rm dim}\, \CH_{BPS}(Y,\beta,j_L,j_R).
\]
The unrefined invariants are BPS indices, 
\[
N(Y,\beta,j_L) = \sum_{j_R\in {1\over 2}\IZ} 
(-1)^{2j_R+1} (2j_R+1)N(Y,\beta,j_L,j_R).
\]
 
String theory arguments \cite{GV-II} 
imply that BPS states should be identified with 
cohomology classes of moduli spaces of stable pure dimension 
sheaves on $Y$. More specifically, let $\CM(Y,\beta,n)$ be the 
moduli space of slope (semi)stable pure dimension one sheaves $F$ 
on $Y$ with numerical invariants 
\[
\ch_2(F)=\beta, \qquad \chi(F)=n.
\]
Suppose furthermore that $(\beta,n)$ are primitive, such that there are 
no strictly semistable points. If $\CM(Y,\beta,n)$ is smooth, the BPS states are 
in one-to-one correspondence with cohomology classes of the moduli space. 
In this case there is a geometric construction of 
the expected 
$SL(2)_L\times SL(2)_R$ action on the BPS Hilbert space 
if in addition there is a natural map $h:\CM(Y,\beta,n)\to \CB$ to a smooth projective variety $\CB$ whose generic fibers are 
smooth Abelian varieties.  
Then \cite{HST} the action follows from decomposition theorem \cite{BBD}, 
\cite{decompth} 
 and as well as the relative hard Lefschetz theorem \cite{RelLefschetz}. 
In particular the action of the positive roots 
$J^+_{L}$, $J^+_R$ 
should is given by cup product with a relative ample class $\omega_h$, 
respectively the pull back of an ample class $\omega_\CB$ on the base. 
One then obtains a decomposition
\[
H^*(\CM(Y,\beta,n))\simeq \bigoplus_{(j_L,j_R)\in \IZ^2} 
R(j_L,j_R)^{\oplus d(j_L,j_R)} 
\]
 where $R(j_L,j_R)$ is the irreducible 
representation of $SL(2)_L\times SL(2)_R$ with highest 
weight $(j_L,j_R)$. 
A priori the multiplicities  $d(j_L,j_R)$  should depend on 
$n$ for a fixed curve class 
$\beta$. Since no such dependence is observed in the low 
energy theory, one is lead to further conjecture that the
$d(j_L,j_R)$ are in fact independent of $n$, 
as long as the numerical invariants $(\beta,n)$ are
 primitive. Granting this additional 
conjecture, the refined BPS invariants are given by 
$N(Y,\beta,j_L,j_R)=d(j_L,j_R)$. 
 
 In more general situations no rigorous mathematical construction 
 of a BPS cohomology theory is known. There is however a rigorous construction of unrefined GV invariants via stable pairs \cite{stabpairs-I,stabpairs-III}
 which will be briefly reviewed shortly. 
 It is worth noting that the BPS cohomology theory would 
 have to
 detect the scheme structure and the obstruction theory of the moduli space
 as is the case in 
 \cite{stabpairs-I,stabpairs-III}.

Concrete examples where the moduli space $\CM(Y,\beta,n)$ is smooth 
are usually encountered in local models, in which case $Y$ is a noncompact threefold. 

\begin{exam}\label{locsurfaces} 
Let $S$ be a smooth Fano surface and $Y$ the total space of 
the canonical bundle $K_S$, 
$\pi:Y\to S$ the natural projection ad $\sigma:S\to Y$ the 
zero section. 
Let $\CO_S(1)$ be a very ample 
lie bundle on $S$. For any coherent sheaf $F$ on $Y$ with 
proper support define the Hilbert polynomial of $F$ by 
\[ 
P_F(m) = \chi(F\otimes_Y \pi^*\CO_S(m)).
\]
For sheaves $F$ with one dimensional support, 
\[ 
P_F(m) = m r_F + n_F , \qquad r_F, n_F \in \IZ, \ r_F>0.
\] 
Such a sheaf will be called (semi)stable
if $$r_F n_{F'}\ (\leq) \ r_{F'}n_F$$ for any proper nontrivial 
subsheaf $0\subset F'\subset F$. Note that 
if $(r_F,n_F)$ are coprime, 
any semistable sheaf with numerical invariants $(r_F,n_F)$ 
must be stable. 
Moreover, Lemma 
 \cite[Lemma 7.1]{algCS} proves that any stable 
sheaf $F$ on $Y$ with Hilbert polynomial $P_F(m)=mr_F+n_F$, with $r_F>0$, must be the extension 
by zero, $F=\sigma_*E$, of a stable sheaf $E$ 
on $S$. Furthermore \cite[Lemma 7.2]{algCS} proves that 
${\rm Ext}^2_S(E,E)=0$.

Let $D$ be an effective divisor on $S$ of degree $d>0$ 
with respect to the polarization $\CO_S(1)$. Let $n\in \IZ$ 
such that $(d,n)$ are coprime. 
Let 
$\CM^s(Y,D,n)$ be the moduli space of stable dimension
one 
sheaves $F$ on $Y$ with numerical invariants 
\[ 
\ch_2(F)=\sigma_*(\ch_1(\CO_D)), \qquad \chi(F) = n, 
\qquad n \in \IZ.
\]
and $\CM^s(S,D,n)$ the moduli space of 
stable dimension one sheaves $E$ on $S$ with 
numerical invariants 
\[
\ch_1(E)=\ch_1(\CO_D), \qquad \chi(E)=n.
\]
Then \cite[Lemma 7.1]{algCS} implies that there  is an 
isomorphism $\CM^s(Y,D,n)\simeq 
\CM^s(S,D,n)$. The vanishing result in \cite[Lemma 7.2]{algCS} 
implies that $\CM^s(S,D,n)$ is smooth according to \cite[Thm. 
4.5.1]{huylehn}. 
Moreover there is a well defined morphism 
$\CM^s(S,D,n)\to |D|$ sending $E$ to its determinant
\cite[Prop. 3.0.2]{detmap}.  
Therefore $\CM^s(Y,D,n)$ is smooth projective and 
equipped with a morphism 
 $h:\CM^s(Y,r,d)\to |D|$.   
\end{exam} 

\begin{exam}\label{loccurves} 
Let $X$ be a smooth projective curve and $D$ an effective 
divisor on $X$, possibly trivial. Let $Y$ be the total space 
of the rank two bundle $\CO_X(-D) \oplus K_X(D)$. Note 
that $H_2(Y)\simeq \IZ$ is generated by the class $\sigma$ 
of the 0 section. Let $\CO_X(1)$ be a very ample line bundle 
on $X$. For any dimension one sheaf $F$ on $Y$ with
proper support define 
\[ 
P_F(m) = \chi(F\otimes_Y \pi^*\CO_X(1)) = mr_F + n_F,
\] 
where $\pi:Y\to X$ is the natural projection.  
Then $F$ is (semi)stable if 
\[
r_Fn_{F'} \ (\leq)\ r_{F'}n_F
\]
for any proper nontrivial subsheaf $0\subset F'\subset F$. 

Let $(d,n)$ be a pair of coprime integers, with $d>0$. 
Then there is a quasi-projective moduli space 
$\CM(Y,d,n)$
of stable dimension one sheaves $F$ on $Y$ 
with proper support and numerical invariants 
\[ 
\ch_2(F)= d\sigma, \qquad \chi(F) =n.
\]
Let $\CH^{e}_r(X)$ be the moduli space of rank $r\geq 1$,
degree $e\in \IZ$ stable Hitchin pairs on $X$. 
Then it is easy to prove the following statements. 

$a)$ If $D=0$ and $(d,n)=1$, there is an isomorphism 
\[
\CM(Y,d,n) \simeq \CH^{n+d(g-1)}_d(X)\times \IC.
\]

$b)$ If $D\neq 0$ and $(d,n)=1$, there is an isomorphism 
\[
\CM(Y,d,n) \simeq \CH^{n+d(g-1)}_d(X).
\]
The proof is analogous to the proof of 
\cite[Thm 1.9]{modADHM},  
the details being omitted. 
\end{exam}

As mentioned above unrefined GV numbers can be defined via Donaldson-Thomas \cite{MNOP-I}
or stable pair invariants \cite{stabpairs-I}.
For smooth projective Calabi-Yau threefolds such invariants are defined 
by integration of virtual cycles on a component of the Hilbert 
scheme of curves, respectively the stable pair  moduli 
space. When $Y$ is a non-compact Calabi-Yau threefold as in 
Example \ref{loccurves}, one has to employ equivariant virtual integration as in \cite{BP,OP} because the moduli spaces are non-compact. The torus action used in this construction is  
a fiberwise action on $Y$ with weights $+1$, $-1$ on the 
direct summands $K_X(D)$, $\CO_X(-D)$ leaving the 
zero section pointwise fixed. Compactness of the fixed 
loci in Donaldson-Thomas theory was proven in \cite{OP} 
while in stable pair theory in \cite{modADHM}. 
%From a string theoretic 
%point of view, this has been explained in \cite{Dijkgraaf:%2006um}
%using IIA/M-theory duality. 
Moreover, the equivalence between 
reduced Donaldson-Thomas theory  and stable pair theory 
has been proven in \cite{Hall_curve_counting} for smooth projective 
Calabi-Yau threefolds. 
Certain versions of this result were also proven in 
\cite{Hilb_DT_PT,Toda_DT_PT}.  
For the quasi-projective varieties 
in Example \ref{loccurves} this equivalence follows in principle 
combining the results of 
\cite{BP,OP},  and 
\cite[Section 5]{MPT}.
To explain this briefly, recall that the local GW, respectively 
DT theory of curves has been computed in 
\cite{BP,OP} using degenerations of $Y$ to normal crossing divisors where each component is a rank two bundle over 
$\IP^1$ and each component intersects at most three other 
along common fibers. Therefore in order to prove their equivalence it suffices to prove equivalence of the resulting relative 
local theories. The same strategy will lead to a proof of 
DT/stable pair correspondence using the results 
of  \cite[Section 5]{MPT} to prove the equivalence of relative 
GW and stable pair theories. The details have not been fully worked out anywhere 
in the literature, but this result will be assumed in this paper. 

In a IIA compactification on $Y$, $\CZ_{DT}(Y,q,Q)$
 is the generating function for the 
degeneracies of BPS states corresponding to bound states of one D6-brane 
and arbitrary D2-D0 brane configurations on $Y$
(see \cite[Sect. 6]{Denef:2007vg}.) 
According to \cite{GV-II, Dijkgraaf:2006um}, M-theory/IIA duality yields an alternative 
expression for this generating function in terms of the five 
dimensional BPS indices
$N(Y,\beta,j_L)$. Then 
\be\label{eq:GVa} 
\CZ_{DT}(Y,q,Q)=\mathrm{exp}\left(F_{GV}(Y,q, Q)\right),
\ee
where 
\be\label{eq:GVb} 
\bal 
F_{GV}(Y,q, Q) = 
\sum_{k\geq 1}\ \sum_{\beta\in H_2(Y),\beta\neq 0}\ 
\sum_{j_L\in {1\over 2}\IZ} 
{Q^{k\beta}\over k} 
(-1)^{2j_L} N(Y,\beta,j_L) {q^{-2kj_L} + \cdots +q^{2kj_L}\over 
(q^{k/2}-q^{-k/2})^{2}}.
\eal 
\ee
Relation \eqref{eq:GVa} can be either inferred from \cite{GV-II} relying 
on the GW/DT correspondence conjectured in \cite{MNOP-I}, or directly derived on physical grounds from Type IIA/M-theory duality \cite{Dijkgraaf:2006um}. Note that the 
generating function $F_{GV}(Y,q,Q)$ in \eqref{eq:GVa} 
may be rewritten in the form \cite{spinBH}
\be\label{eq:GVab}
F_{GV}(Y,q,Q) = \sum_{g\geq 0} \sum_{\beta\neq 0} 
n_{g,\beta} u^{2g-2} \sum_{k\geq 1} 
{1\over k} \left({\mathrm{sin}(ku/2)\over \mathrm{u/2}}
\right)^{2g-2} Q^{k\beta} 
\ee
where $q= -e^{-iu}$ and 
\[
N(Y,\beta,j_L) = \sum_{g\geq 2j_L} \binom{2g}{q+2j_L}n_{g,\beta}.
\]
In the mathematics literature relation \eqref{eq:GVa}
with $F_{GV}(Y,q,Q)$ of the form \eqref{eq:GVab}, 
where $n_{g,\beta}\in \IZ$ is known as the strong 
rationality conjecture \cite{stabpairs-I}. 
It was proven for irreducible curve classes 
on smooth projective Calabi-Yau threefolds in 
\cite{stabpairs-III} 
and for general curve classes in \cite{generating_wallcrossing} 
with a technical caveat  
concerning holomorphic Chern-Simons functions for 
perverse coherent sheaves. In all these cases the proof
does not provide a cohomological interpretation of the invariants 
$n_{g,\beta}$. 

According to \cite{Iqbal:2007ii}, a similar relation is expected to hold 
between refined stable pair invariants and the GV numbers 
$N(Y,\beta, j_L,j_R)$. 
As explained in \cite{DG} refined stable pair invariants are obtained as a 
specialization of the virtual motivic invariants of Kontsevich and Soibelman 
\cite{wallcrossing}. 
Then one expects \cite{Iqbal:2007ii} a relation of the form 
\be\label{eq:refGVa}
\CZ_{DT,Y}(q,Q,y) = \mathrm{exp}\left(F_{GV,Y}(q, Q,y)\right),
\ee
where 
\be\label{eq:refGVb} 
\bal 
F_{GV,Y}(q,Q,y) = & 
\sum_{k\geq 1}\ \sum_{\beta\in H_2(Y),\beta\neq 0}\ 
\sum_{j_L,j_R\in {1\over 2}\IZ} 
{Q^{k\beta}\over k} 
(-1)^{2j_L+2j_R} N(Y,\beta,{j_L,j_R}) \\
& 
q^{-k} {(q^{-2kj_L} + \cdots +q^{2kj_L})(
y^{-2kj_R}+\cdots+y^{2kj_R})
\over 
(1-(q y)^{-k})(1-(q y^{-1})^{-k})}.\\
\eal 
\ee
The 
expression \eqref{eq:refGVb} was written in \cite{Iqbal:2007ii} in different variables, $(q^{-1}y, q^{-1}y^{-1})$. 

The main goal of this note is to point out that the refined 
GV expansion \eqref{eq:refGVa} for a local 
curve geometry is related via a simple change of 
variables to the Hausel-Rodriguez-Villegas formula for 
character varieties. There a few conjectural steps involved 
in this identification. First, it relies on a explicit conjectural formula 
for the refined stable pair theory of a local curve 
derived in section (\ref{three}) from geometric engineering and 
instanton sums. In fact, it is expected that a rigorous construction 
of motivic stable pair theory of local curves should be possible 
following the program of Kontsevich and Soibelman 
\cite{wallcrossing}. A conjectural motivic formula generalizing 
equation \eqref{eq:refinvB} 
has been recently written down by Mozgovoy \cite{adhm-rec}. 
Second, as explained in detail in section (\ref{four}), the 
refined GV invariants of the local curve are in fact perverse 
Betti numbers of the Hitchin moduli space. Therefore, the 
conversion of the HRV formula into a refined GV expansion 
relies on the identification between the weight filtration 
on the cohomology of character varieties and the perverse 
filtration on the cohomology of the Hitchin system 
conjectured by de Cataldo, Hausel and Migliorini 
\cite{hodge-char}. This will be referred to as the $P=W$ 
conjecture. The connection
found here provides independent physics based evidence for this conjecture. 
Finally, note that further evidence for all the claims of the present paper comes from the 
recent 
rigorous results of 
\cite{adhm-rec,MY,MS}.  In \cite{adhm-rec} it is 
rigorously proven that the refined theory of the local curve 
implies the HRV conjecture for the Poincar\'e polynomial 
of the Hitchin system via motivic wallcrossing while \cite{MY,MS} prove 
expansion formulas analogous to \eqref{eq:refGVa} for families 
of irreducible reduced plane curves. 

{\it Acknowledgements.} We are very grateful to Tamas Hausel and 
Fernando Rodriguez-Villegas for illuminating discussions on their 
work. We would also like to thank Jim Bryan, Ugo Bruzzo, Ron Donagi,
Oscar Garcia-Prada, Lothar G{\"o}ttsche, Jochen Heinloth, Dominic Joyce,
Ludmil Katzarkov, Bumsig Kim, Melissa Liu, Davesh Maulik, Greg Moore, Sergey
Mozgovoy, 
Kentaro Nagao,
Alexei Oblomkov, Rahul Pandharipande, Tony Pantev, Vivek Shende, Artan Sheshmani, 
Alexander 
Schmitt, Jacopo Stoppa, Balazs Szendroi, Andras Szenes, Michael Thaddeus,
Richard Thomas, and  Zhiwei Yun for very helpful 
conversations.  D.-E.D. would like to thank the organizers of the 
Moduli Space Program 2011 at Isaac Newton Institute for the partial 
support during completion of this work, as well as a very stimulating 
mathematical environment. The work of D.-E.D. was also supported
in part by NSF grant PHY-0854757-2009. 
The work of W.-Y.C. was supported by NSC grant 99-2115-M-002-013-MY2 
and Golden-Jade fellowship of Kenda Foundation.

\section{Hausel-Rodriguez-Villegas formula and $P=W$}\label{two}

Let $X$ be a smooth projective curve over $\IC$ of genus $g\geq 1$, and $p\in X$ an arbitrary closed point. Let $\gamma_p
\in \pi_1(X\setminus \{p\})$ be the natural generator associated 
to $p$. 
For any coprime integers $r\in \IZ_{\geq 1}$, $e\in \IZ$, 
the character variety $\CC_r^e(X)$ is the moduli space of 
representations 
\[
\phi:\pi_1(X\setminus\{p\}) \to GL(r,\IC),\qquad 
\phi(\gamma_p) = e^{2i\pi e/r} I_r
\]
modulo conjugation. $\CC_r^e(X)$ is a smooth quasi-projective 
variety, and its rational cohomology $H^*(\CC_r^e(X))$ carries a weight filtration 
\be\label{eq:mixedhodgeA} 
W^k_0 \subset \cdots W^k_i \subset \cdots \subset W^k_{2k} = H^k(\CC^n_r(X)).
\ee
According to \cite{HRV}, $W^k_{2i}=W^k_{2i+1}$ for all $i=0,\ldots, 2k$, hence one can define  
the  mixed Poincar\'e polynomial 
\be\label{eq:virtpoincareA}
W(\CC_r^e(X),z,t) = \sum_{i,k}\mathrm{dim}(W^k_i/W^k_{i-1}) t^k z^{i/2}. 
\ee
Moreover it was proven in \cite{HRV} 
that $W(\CC_r^e(X),z,t)$ 
is independent of $e$ for fixed $r$, with $(r,e)$ coprime. Therefore it will 
be denoted below by $W_r(z,t)$. 
Obviously  $W_r(1,t)$ is the usual Poincar\'e polynomial. Note that one can equally well use compactly supported 
cohomology in \eqref{eq:virtpoincareA}, which is related to
cohomology without support condition by Poincar\'e 
duality \cite{HRV},  
\[
H_c^k(\CC^n_r(X)) \times H^{2d-k}(\CC^n_r(X)) \to \IC.
\]
The difference would be an irrelevant 
overall monomial factor. 
Using number theoretic considerations Hausel-Rodriguez-Villegas 
\cite{HRV} derive a conjectural formula for the mixed Poincar\'e polynomials $W_r(z,t)$ as follows. 

\subsection{Hausel-Rodriguez-Villegas formula}\label{HRVsect}
The conjecture formulated in \cite{HRV} expresses the
generating function 
\[
F_{HRV}(z,t,T) =
\sum_{r,k\geq 1} B_r(z^k,t^k) W_r(z^k,t^{k}) {T^{kr}\over k},
\]
\[
B_r(z,t) = {(zt^2)^{(1-g)r(r-1)}\over (1-z)(1-zt^2)},
\]
as 
\be\label{eq:HRVa}
F_{HRV}(z,t,T) ={\rm{ln}}\, Z_{HRV}(z,t,T)
\ee
where $Z_{HRV}(z,t,T)$ is a sum of rational functions associated 
to Young diagrams. 
Given a Young diagram $\mu$ as shown below
\bigskip
\bigskip 

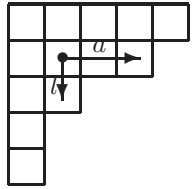
\setlength{\unitlength}{0.8mm}
\begin{picture}(150,40) 
\thicklines
\put(20,40){\line(1,0){30}}
\put(20,34){\line(1,0){30}}
\put(20,28){\line(1,0){24}}
\put(20,22){\line(1,0){12}} 
\put(20,16){\line(1,0){6}}
\put(20,10){\line(1,0){6}}
\put(20,40){\line(0,-1){30}}
\put(26,40){\line(0,-1){30}}
\put(32,40){\line(0,-1){18}}
\put(38,40){\line(0,-1){12}}
\put(44,40){\line(0,-1){12}}
\put(50,40){\line(0,-1){6}}
\put(28,30){$\bullet$}
\put(29,31){\vector(1,0){13}}
\put(29,31){\vector(0,-1){7}}
\put(34,32){$a$}
\put(27,25){$l$}
%\put(70,40){{\huge $Z_{HRV}=\sum_{\mu} T^{|\mu|} \CH_\mu(z,t)      
 %$}}
%\put(70,25){ $|\mu|=\sharp \mathrm{boxes\ of} \ \mu$}
%\put(70,10){ $h(\Box)=a(\Box)+l(\Box)+1$}
\end{picture}

\noindent
let $\mu_i$ be the length of the $i$-th 
row, $|\mu|$ the total number of boxes of $\mu$, 
and $\mu^t$ the transpose of $\mu$. 
For any box $\Box=(i,j)\in \mu$ let 
\[
a(\Box)=\mu_i-j,\qquad l(\Box)=\mu^t_j-i,\qquad 
h(\Box)=a(\Box)+l(\Box)+1,
\] 
be the arm, leg, respectively hook length.
Then 
\[
\CZ_{HRV}(z,t,T)=\sum_{{\mu}} \CH_{g}^\mu(z,t) T^{|\mu|}
\]
where
\[
\CH_g^\mu(z,t)=
\prod_{\Box \in \mu} 
\frac{(zt^2)^{l(\Box)(2-2g)}(1-z^{h(\Box)}t^{2l(\Box)+1})^{2g}}{(1-z^{h(\Box)}t^{2l(\Box)+2})
(1-z^{h(\Box)}t^{2l(\Box)})}.
\]
The main observation in this note is that equation \eqref{eq:HRVa} 
can be identified with the expansion of the refined Donaldson-Thomas
series of a certain Calabi-Yau threefold in terms of numbers of BPS
states. 

\subsection{Hitchin system and $P=W$}\label{PWsection}
Let $\CH^e_r(X)$ be the moduli space of stable Higgs bundles 
$(E,\Phi)$ on $X$, where $\Phi$ is a Higgs field with coefficients 
in $K_X$. For coprime $(r,e)$ this is a smooth quasi-projective 
variety equipped with a projective Hitchin map 
\[ 
h:  \CH^e_r(X)\to \CB
\]
to the affine variety 
\[
\CB = \oplus_{i=1}^r H^0(K_X^{\otimes i}).
\]
The decomposition of the derived 
direct image $Rh_*{\underline \IQ}$ 
into perverse sheaves yields \cite{decompth,hodge-char}
a perverse filtration 
\[
0=P^k_0 \subset P^k_1\subset \cdots \subset P^k_k=H^k(\CH^e_r(X))
\]
on cohomology. Following the construction in 
\cite[Sect. 1.4.1]{hodge-char}, let $H^k(\CB, Rh_*{\underline \IQ})$ 
denote the $k$-th hypercohomology group and 
${}^{\mathfrak p}\tau_{\leq p} Rh_*{\underline \IQ}$ 
denote the truncations of $Rh_*{\underline \IQ}$. 
Then set 
\[
P_pH^k(\CB, Rh_*{\underline \IQ}) 
= {\rm Im}\big( H^k(\CB, {}^{\mathfrak p}\tau_{\leq p} Rh_*{\underline \IQ}) \to H^k(\CB, 
{}^{\mathfrak p}\tau_{\leq p} Rh_*{\underline \IQ})\big).
\]
and 
\[
P^k_p = P_pH^{k-d}({\CB}, Rh_*{\underline \IQ}[d]) 
\]
where $d={\rm dim}_\IC \CB$. 

It is well known that $\CC^e_r(X)$ and $\CH^e_r(X)$ are 
identical as smooth real manifolds. 
%More precisely their complex 
%structures are related by a hyper-K\"ahler rotation. 
This result is due to \cite{hitchin-selfd,twistedharmonic} for 
rank $r=2$ and \cite{flatGbundles,Higgslocal} for general $r\geq 2$. 
Therefore there is a natural identification 
$H^*(\CC^e_r(X))=H^*(\CH^e_r(X))$.
 Then it is conjectured in  \cite{hodge-char} that the two filtrations 
 $W^k_j$, $P^k_j$ coincide, 
 \[
 W^k_{2j}=P^k_j
 \]
 for all $k,j$. This is proven in \cite{hodge-char}
 for Hitchin systems of rank $r=2$. 
 
For future reference note that a relative ample class $\omega$ 
with respect to $h$
yields a hard Lefschetz isomorphism  \cite{RelLefschetz}
\[
\omega^l : Gr^P_{d-l}H^k(\CH^e_r(X)) {\buildrel \sim \over \longrightarrow} Gr^P_{d+l}H^{k+2l}(\CH^e_r(X)).
\]
This is known under the name of relative hard Lefschetz theorem.

Note that granting the $P=W$ conjecture, equation \eqref{eq:HRVa} yields explicit formulas for the perverse 
Poincar\'e polynomial of the Hitchin moduli space. In particular, by specialization to $z=1$ it determines the Poincar\'e polynomial of the Hitchin moduli space of any rank $r\geq 1$.

\section{Refined stable pair invariants of local curves}\label{three}

Let $Y$ be the total space of the rank two bundle $\CO_X(-D)\oplus 
K_X(D)$ where $D$ is an effective divisor of degree $p\geq 0$ 
on $X$ as in Example (\ref{loccurves}). Note that $H_2(Y)\simeq \IZ$ is generated by the class 
$\sigma$
of the zero section. 
Following \cite{stabpairs-I}, 
stable pairs on $Y$ are two term complexes 
$P=(\CO_Y {\buildrel s\over \longto} F)$ where $F$ is a pure dimension one sheaf and 
$s$ a generically surjective section. Since $Y$ is noncompact, in the 
present case, it will be also required that $F$ have compact support,
which must be necessarily a finite cover of $X$. The numerical invariants 
of $F$ will be 
\[
\ch_2(F) =d\sigma, \qquad \chi(F)=n.
\]
Then according to \cite{stabpairs-I}, there is a quasi-projective fine 
moduli space $\calP(Y,d,n)$ of pairs of type $(d,n)$
equipped with a symmetric 
perfect obstruction theory. 
The moduli space also carries a torus action induced by the
$\IC^\times$ action on $Y$ which scales $\CO_X(-D),K_X(D)$ with weights 
$-1,1$. 
Virtual numbers of stable pairs can be defined by 
equivariant virtual integration by analogy with 
\cite{BP,OP}. On smooth projective Calabi-Yau threefolds,  the virtual number of pairs is equal to the
 Euler characteristic of the 
moduli space weighted by the Behrend function 
 \cite{micro}. 
The analogous relation, 
\[
P(\beta,n) = \chi^B(\calP(Y,d,n)),
\]
for equivariant residual invariants of local curves follows from 
\cite[Thm. 1.9]{modADHM} and  \cite[Lemma 3.1]{chamberII}. 
Let 
\[
Z_{PT}(Y,q,Q) = 1+ \sum_{d\geq 1}\sum_{n\in \IZ} P(d,n)Q^dq^n.
\]
Applying the motivic Donaldson-Thomas formalism of Kontsevich and Soibelman, one obtains a refinement $P^{ref}(d,n,y)$ of stable pair invariants
modulo foundational issues. The $P^{ref}(d,n,y)$ are Laurent polynomials of the formal variable $y$ with integral coefficients. 
In a string theory compactification on $Y$ these coefficients are 
numbers of D6-D2-D0 bound states with given four dimensional 
spin quantum number. 
The resulting generating series will 
be denoted by $Z_{PT}^{ref}(Y,q,Q,y)$. 

\subsection{TQFT formalism}

A TQFT formalism for unrefined Donaldson-Thomas theory of a local 
curve has been developed in \cite{OP}, in parallel with a similar 
construction \cite{BP} in Gromov-Witten theory. 
Very briefly, the final result is that the generating series of local invariants is obtained by gluing vertices corresponding 
to a pair of pants decomposition of the Riemann surface $X$. 
Each such vertex is a rational function $P_{\mu_i}(q)$
 labelled by three 
partitions $\mu_i$, $i=1,2,3$ corresponding to the three boundary 
components. In the equivariant Calabi-Yau case a nontrivial result is 
obtained only for identical partitions, $\mu_i=\mu$, $i=1,2,3$, in which 
case 
\[
P_\mu(q)=  \prod_{\Box\in \mu} 
(q^{h(\Box)/2}-q^{-h(\Box)/2}).
\]
Then the generating function is given by 
\be\label{eq:tqftA} 
Z_{DT}(Y,q,Q) = \sum_{\mu} (-1)^{p|\mu|} 
q^{-(g-1-p)\kappa(\mu)}(P_\mu(q))^{2g-2} Q^{|\mu|}
\ee
where 
\[
\kappa(\mu) = \sum_{\Box\in \mu}(i(\Box)-j(\Box)).
\]

\subsection{Refined invariants from instanton sums}

Although the refined stable pair invariants are not rigorously 
constructed for higher genus local curves, 
string duality leads to an explicit conjectural 
formula for the series $Z_{PT}^{ref}(Y,q,Q,y)$. This follows 
using geometric engineering 
\cite{geom_eng,fivedone,fivedtwo,Lawrence:1997jr} of 
supersymmetric five 
dimensional gauge theories. 

For completeness, geometric 
engineering is a correspondence between local Calabi-Yau 
threefolds and five dimensional gauge theories 
with eight supercharges. Such gauge theories are classified 
by triples $(G,R,p)$, where $G$ is a compact semisimple Lie 
group and $R$ a unitary  representation of $G$, and $p\in \IZ$. 
Physically $R$ encodes the matter content of the theory, 
and $p$ is the level of a five dimensional Chern-Simons term.
For certain triples $(G,R,p)$ (but not all) 
there exists a noncompact smooth Calabi-Yau 
threefold $Y_{(G,R,p)}$ such that the 
gauge theory specified 
by $(G,R,p)$ is the extreme infrared limit of M-theory 
in the presence of a gravitational background 
specified by $Y_{(G,R,p)}$. 
Many such examples are known \cite{geom_eng,KMV}, 
but the list is not exhaustive, and there is no known 
necessary and sufficient condition on $(G,R,p)$ 
guaranteeing the 
existence of $Y_{(G,R)}$. 

For example if $G=SU(N)$, $N\geq 2$, $R$ is 
the zero 
representation, and $p=0$, the corresponding threefold $Y_{(SU(N),0,0)}$ is 
constructed as follows. 
Let $Y$ be the total space of the rank two bundle
$\CO_{\IP^1}\oplus \CO_{\IP^1}(-2)$ and let $\mu_N$ be the 
multiplicative group of $N$-th roots of unity. 
There is a fiberwise action $\mu_N\times 
Y\to Y$ where the generator $\eta = e^{2i\pi/N}$ 
acts by multiplication by $(\eta, \eta^{-1})$ on the two 
summands. The quotient $Y/\mu_N$ is a singular toric variety. 
Then $Y_{(SU(N),0,0)}$ is the unique toric crepant resolution 
of $Y/\mu_N$.  

Moreover, suppose $Y$ is replaced in previous paragraph 
by a rank two bundle of the form  $\CO_X(-D)
\oplus \CO_X(K_X+D)$, 
with $X$ a curve of genus $g\geq 1$, as in Example
 \ref{loccurves}. Then there exists a corresponding gauge theory, and it has gauge group $G=SU(N)$, matter content   
$R = ad(G)^{\oplus g}$, and level $p={\rm deg}(D)$, 
where $ad(G)$ denotes the 
adjoint representation. 

An important mathematical prediction of this correspondence
is  an identification between
a generating function of stable pair invariants 
of $Y_{(G,R,p)}$ and the five dimensional equivariant instanton 
sum of the gauge theory $(G,R,p)$ 
defined by Nekrasov in \cite{Nekrasov:2002qd}.
Some care is needed in formulating a precise relation; 
since $Y_{(G,R,p)}$ are noncompact, the stable pair invariants must be defined as  residual equivariant 
invariants  with respect to a torus action. 
In addition, this identification also involves a nontrivial change of 
formal variables which is known in many examples, 
but has no general prescription.  

Therefore a more precise formulation of this conjecture 
would state that there exists a torus action on 
$Y_{(G,R,p)}$ such that the residual equivariant
stable pair theory is well defined, and its generating 
function equals the equivariant instanton sum of the
gauge theory $(G,R,p)$ up to change of variables. 
Such statements have been formulated and 
proved in many examples where 
$Y_{(G,R,p)}$ is a toric Calabi-Yau threefold in  
\cite{EK-I,IKP-I,IKP-II,EK-II,Hollowood:2003cv,Konishi-I,LLZ,Iqbal:2007ii}. 
Furthermore, a refined version of the geometric engineering 
conjecture is available due to the work of \cite{IKV}, 
where it has been checked for  
$SU(N)$ with $N=2,3$, and $(R,p)=(0,0)$. 

In the present case, the geometric engineering 
conjecture yields \cite{wall-pairs} an explicit prediction 
for the residual stable pair theory of the threefolds $Y$ 
in Example \ref{loccurves}. Because of a subtlety of physical 
nature, this case was treated in \cite{wall-pairs}  as a limit of 
$SU(2)$ gauge theory with $R=ad(G)^{\oplus g}$
and level $p={\rm deg}(D)$.
Omitting the computations, which are given in detail 
in \cite[Sect 3]{wall-pairs}, 
note that the final result can be presented in terms
of quivariant K-theoretic invariants of the Hilbert scheme of 
points in $\IC^2$ as follows. 

Let ${\mathcal Hilb}^k(\IC^2)$ denote the Hilbert scheme 
of length $k\geq 1$ zero dimensional subschemes of 
$\IC^2$. It is smooth, quasi-projective and carries a  
${\bf G}={\IC^\times}\times 
\IC^\times$-action 
 induced by the natural scaling action on $\IC^2$. 
 Let   $R_{\bf G}$ denote the representation ring of 
${\bf G}$, and $q_1,q_2:{\bf G}\to \IC$ the characters defined by 
\[
q(t_1,t_2)=t_1, \qquad q_2(t_1,t_2) =t_2.
\]
Let also 
$\ch : R_{\bf G}\to \IZ[q_1,q_2]$ 
denote the canonical ring isomorphism assigning to any representation $R$ the character $\ch(R)$. 

Now let 
tautological vector bundle $\CV_k$ on the Hilbert 
scheme whose fiber at a point 
$[Z]$ is the space of global sections $H^0(\CO_Z)$. 
For each pair of integers $(g,p)\in \IZ^2$, $g\geq 0$, 
$p\geq 0$  let 
\[
\CE_k^{g,p} = 
T^*{\mathcal Hilb}^k(\IC^2)^{\oplus g} \otimes 
\det(\CV_k)^{1-g-p}.
\]
By construction, $\CE_k^{g,p}$ has a natural ${\bf G}$-equivariant structure
which yields a linear {\bf G}-action on the sheaf cohomology groups $H^i(\Lambda^j\CE_k^{g,p})$ of its exterior 
powers. Moreover, as observed for example 
in \cite{LLZ} although these spaces 
are infinite dimensional, each irreducible representation 
of ${\bf G}$ has finite multiplicity in the decomposition of  $H^i(\Lambda^j\CE_k^{g,p})$. 
Therefore one can formally define the equivariant 
$\chi_{\tilde y}$-genus of $\CE_k^{(g,p)}$, 
\[
\chi_{\widetilde y} ( {\mathcal E}_k^{(g,p)}) = \sum_{i,j}
(-{\widetilde y}\, )^j(-1)^i {\mathrm{\huge ch}}\, H^i(\Lambda^j {\CE}_k^{(g,p)}) \\
\]
as an element of $\IZ[[q_1,q_2]]$. 
The equivariant K-theoretic partition function 
is defined by 
\be\label{eq:instsumA} 
\begin{aligned} 
 Z_{inst}(q_1,q_2,\, {\widetilde Q},\, {\widetilde y}) = 
 \sum_{k\geq 0} 
\chi_{\widetilde y}(\CE_k^{g,p}) {\widetilde Q}^k.\\
\eal
\ee
A fixed point theorem gives an explicit formula 
for 
$Z_{inst}(q_1,q_2,\, {\widetilde Q},\, {\widetilde y})$ 
as a sum over partitions: 
\[
\begin{aligned} 
& Z_{inst}(q_1,q_2,{\widetilde Q},{\widetilde y}\, ) =
\sum_{\mu} 
\prod_{\Box\in \mu} (q_1^{-l(\Box)} q_2^{-a(\Box)})^{g-1+p} 
\\
&
{(1-{\widetilde y} q_1^{-l(\Box)} q_2^{a(\Box)+1})^g 
 (1-{\widetilde y} q_1^{l(\Box)+1} q_2^{-a(\Box)})^g 
 \over (1- q_1^{-l(\Box)} q_2^{a(\Box)+1})
 (1- q_1^{l(\Box)+1} q_2^{-a(\Box)})} {\widetilde Q}^{|\mu|}
\end{aligned} 
\]
The resulting conjectural expression for the refined stable pair
partition function is then \cite{wall-pairs}
\be\label{eq:refinvA}
Z^{ref}_{PT}(Y,q,Q,y) = 
Z_{inst}(q^{-1}y,qy,(-1)^{g-1}y^{2-g}Q,y^{-1}).
\ee
A straightforward computation shows that 
\be\label{eq:refinvB}
Z^{ref}_{PT}(Y,q,Q,y) = \sum_{\mu}
\Omega^\mu_{g,p}(q,y) Q^{|\mu|}
\ee
where
\[
\bal 
\Omega^\mu_g(q,y) = & \ (-1)^{p|\mu|}
\prod_{\Box\in \mu}\bigg[ \big(q^{l(\Box)-a(\Box)}y^{-(l(\Box)+a(\Box))}\big)^p (qy^{-1})^{(2l(\Box)+1)(g-1)} \\
&\ \ \ 
{(1-q^{-h(\Box)}y^{l(\Box)-a(\Box)})^{2g}\over 
(1-q^{-h(\Box)} y^{l(\Box)-a(\Box)-1})
(1-q^{-h(\Box)} y^{l(\Box)-a(\Box)+1})}\bigg].\\
\eal
\]
The change of variables in \eqref{eq:refinvA} 
does not have a conceptual 
derivation. This conjecture is supported by extensive numerical 
computations involving wallcrossing for refined invariants in \cite{wall-pairs}. 
Further supporting evidence for the formula \eqref{eq:refinvA} 
is obtained by comparison with the unrefined TQFT formula 
\eqref{eq:tqftA} for local curves. Specializing the right 
hand side of \eqref{eq:refinvA} at $y=1$, one obtains 
\[ 
Z^{ref}_{PT}(Y,q,Q,1) = \sum_{\mu} Q^{|\mu|}\prod_{\Box\in \mu} 
(-1)^{p|\mu|}q^{(g-1+p)(l(\Box)-a(\Box))} (q^{h(\Box)/2}-q^{-h(\Box)/2})^{2g-2}.
\]
Agreement with \eqref{eq:tqftA} follows from the identity 
\[ 
\sum_{\Box\in \mu}(l(\Box)-a(\Box)) = \sum_{\Box\in 
\mu} (j(\Box) - i(\Box)) = -\kappa(\mu). 
\]

Finally, note that the expression \eqref{eq:refinvA} with $p=0$
is related to 
the left hand side of the HRV formula by 
\be\label{eq:HRVGVa}
Z_{HRV}(z,t,T) = Z_{PT}^{ref}(Y,(zt)^{-1},(zt^2)^{g-1}T,t).
\ee

\section{HRV formula as a refined GV expansion}\label{four}

This section spells out in detail the construction of refined 
GV invariants of a threefold $Y$ as in Example 
\ref{loccurves}  with $p={\rm deg}(D)=0$ 
in terms of the perverse filtration on the 
cohomology of the Hitchin moduli space. 
In this case the generic 
fibers and the base of the Hitchin map 
$h: \CH^e_r(X)\to \CB$ have equal complex dimension $d$. 
Using the conjectural formula
\eqref{eq:refinvA}, it will be shown that equation \eqref{eq:refGVa} yields the HRV formula 
by a monomial change of variables. 
As observed in Example \ref{loccurves}, 
the moduli space of slope stable pure dimension 
one sheaves $F$ on $Y$ with compact support 
 and numerical invariants 
\[
\ch_2(F) = r\sigma, \qquad \chi(F)=n 
\]
 is isomorphic to $\IC\times \CH_r^{n+r(g-1)}(X)$ provided that 
 $(r,n)=1$.  
Therefore, following the general arguments in the introduction, 
one should be able to define refined GV invariants using the 
decomposition theorem for the Hitchin map 
$h: \CH_r^{e}(X)\to \CB$, $e=n+r(g-1)$. 
However, since the base of the Hitchin fibration is a linear space, there will 
not exist an $SL(2)_L\times SL(2)_R$ action on cohomology as 
required by M-theory. 
In this situation one can only define 
an $SL(2)_L\times \IC^\times_R$-action 
where $\IC^\times_R$ can be thought of as a Cartan 
subgroup of $SL(2)_R$. 
This action can be explicitly described in terms of the perverse sheaf filtration 
constructed in \cite[Sect. 1.4]{hodge-char}, which was 
briefly reviewed in Section \ref{PWsection}.

Note that given a
relative ample class $\omega$ for the Hitchin map
there is a preferred splitting 
\be\label{eq:prefsplitting}
H^k(\CH^e_r(X)) \simeq \bigoplus_{p} Gr_pH^k(\CH^e_r(X))
\ee
of the perverse sheaf 
filtration presented in detail in 
\cite[Sect 1.4.2, 1.4.3]{hodge-char}.
Moreover, the relative Lefschetz isomorphism 
\[
\omega^l : Gr^P_{d-l}H^k(\CH^e_r(X)) {\buildrel \sim \over \longrightarrow} Gr^P_{d+l}H^{k+2l}(\CH^e_r(X)).
\]
yields a decomposition 
\[
 Gr^P_pH^k(\CH^e_r(X)) \simeq \bigoplus_{i+2j=p} Q^{i,j;k}, \qquad 
Q^{i,j;k}=\omega_h^j Q^{i,0;k-2j}.
\]
where 
\[
Q^{i,0;k} = {\rm Ker}\big(\omega_h^{d-i+1}:
Gr^P_{i}H^k(\CH^e_r(X))\to Gr^P_{2d-i+1}H^{k+2(d-i+1)}(\CH^e_r(X))\big)
\]
for all $0\leq i \leq d$. 
Let $Q^{i,j}=\bigoplus_{k\geq 0} Q^{i,j;k}$. 
By construction, for fixed $0\leq i\leq d$, there is an isomorphism 
\[
\bigoplus_{j=0}^{d-i} Q^{i,j} \simeq  R_{(d-i)/2}^{\oplus \mathrm{dim}(Q^{i,0}) }
\]
where $R_{j_L}$ is the irreducible representation of $SL(2)_L$ with spin 
$j_L\in {1\over 2}\IZ$. The generator $J_L^+$ is represented by cup-product with 
$\omega$, and  $Q^{i,j}$ is the eigenspace of the Cartan generator $J_L^3$ with eigenvalue
$j-(d-i)/2$. Note that cup-product with $\omega$ preserves the grading 
$k-d-2j$ therefore one can define an extra $\IC^\times$-action on $H^*(\CH_r^e(X))$ which scales $Q^{i,j;k}$ with weight $d+2j-k$. This torus action will be denoted by 
$\IC^\times_R \times H^*(\CH_r^e(X))\to H^*(\CH_r^e(X))$.
Note also that 
\[
d+ 2j-k \geq -d 
\]
since $j\geq 0$ and $k\leq -2d$. 

In conclusion, in the present local curve geometry the $SL(2)_L\times SL(2)_R$ action on the cohomology of the moduli space of D2-D0 branes 
is replaced by an $SL(2)_L\times \IC^\times_R$ action. This is 
is certainly puzzling from a physical perspective 
since the BPS states are expected to form five-dimensional spin multiplets. 
The absence of a manifest $SL(2)_R$ symmetry of the local BPS spectrum 
is due to noncompactness of the moduli space. 
This is simply a symptom of the fact that there is no well defined physical
decoupling limit 
associated to a local higher genus curve as considered here in M-theory. 
In principle, in order to obtain a physically sensible theory, one would have to construct 
a Calabi-Yau threefold ${\overline Y}$ containing a curve $X$ with 
infinitesimal neighborhood isomorphic to $Y$
so that the moduli space $\CM_{\overline Y}(r[X],n)$ is compact
and there is an  embedding $H^*(\CM_Y(r,n))\subset H^*(\CM_{\overline Y}(r[X],n))$. 
The cohomology classes in the complement would then provide the missing components of the five-dimensional spin multiplets. 
Such a construction seems to be very difficult, and it is not in fact needed for the 
purpose of the present paper.

Given the $SL(2)_L\times \IC^\times_R$ action 
described in the previous paragraph, one can define the following local version 
of the refined Gopakumar-Vafa expansion 
\eqref{eq:refGVb}. 
\be\label{eq:refGVc} 
\bal 
F_{GV,Y}(q,Q,y) = & 
\sum_{k\geq 1}\sum_{r\geq 1}
\sum_{j_L=0}^{d/2}\sum_{l\geq -d} 
{Q^{kr}\over k} 
(-1)^{2j_L+l} N_r({(j_L,l)})\\
& \qquad\qquad 
{q^{-k} (q^{-2kj_L} + \cdots +q^{2kj_L})
y^{kl}\over 
(1-(q y)^{-k})(1-(q y^{-1})^{-k})}.\\
\eal 
\ee
where 
\[
N_r({j_L,l})= \mathrm{dim}(Q^{d-2j_L,0;d+l}).
\] 
The same change of variables as in equation 
\eqref{eq:HRVGVa} yields 
\be\label{eq:HRVGVb} 
\bal
F_{GV,Y}((zt)^{-1},(zt^2)^{g-1}T,t) = & 
\sum_{k\geq 1}\sum_{r\geq 1}
{T^{kr}\over k} 
B_r(z^k,t^k) 
P_r(z^k, t^k)
\eal \ee
where $B_r(z,t)$ is defined above equation \eqref{eq:HRVa}, 
and  
\[
\bal 
P_r(z,t)= \sum_{j=0}^{d} 
\sum_{l\geq 0} 
(-1)^{j+l} N_r((j-d)/2,l-d)\, 
t^l\, (1 + \cdots +(zt)^{2j}).\\
\eal 
\]
Now it is clear that the change of variables 
\[ 
(q,Q,y) = ((zt)^{-1},(zt^2)^{g-1}T, t)
\]
identifies the 
HRV formula \eqref{eq:HRVa} 
with the refined GV expansion \eqref{eq:refGVa} 
for a local curve provided that 
\be\label{eq:HRVGVc} 
P_r(z,t) = W_r(z,t). 
\ee
However, given the cohomological definition of the refined GV
invariants $N_r(j_L,l)$, relation \eqref{eq:HRVGVb} 
follows from the $P=W$ conjecture of \cite{hodge-char}. 
This provides a string theoretic explanation as well as strong 
evidence for this conjecture.

\bibliography{adhmref.bib}

\begin{thebibliography}{10}

\bibitem{micro}
K.~Behrend.
\newblock Donaldson-{T}homas type invariants via microlocal geometry.
\newblock {\em Ann. of Math. (2)}, 170(3):1307--1338, 2009.

\bibitem{BBD}
A.~A. Be{\u\i}linson, J.~Bernstein, and P.~Deligne.
\newblock Faisceaux pervers.
\newblock In {\em Analysis and topology on singular spaces, {I} ({L}uminy,
  1981)}, volume 100 of {\em Ast\'erisque}, pages 5--171. Soc. Math. France,
  Paris, 1982.

\bibitem{Hall_curve_counting}
T.~Bridgeland.
\newblock Hall algebras and curve-counting invariants.
\newblock {\em J. Amer. Math. Soc.}, 24(4):969--998, 2011.

\bibitem{BP}
J.~Bryan and R.~Pandharipande.
\newblock The local {G}romov-{W}itten theory of curves.
\newblock {\em J. Amer. Math. Soc.}, 21(1):101--136 (electronic), 2008.
\newblock With an appendix by Bryan, C. Faber, A. Okounkov and Pandharipande.

\bibitem{wall-pairs}
W.-y. Chuang, D.-E. Diaconescu, and G.~Pan.
\newblock {Wallcrossing and Cohomology of The Moduli Space of Hitchin Pairs}.
\newblock {\em Commun.Num.Theor.Phys.}, 5:1--56, 2011.

\bibitem{chamberII}
W.-y. Chuang, D.-E. Diaconescu, and G.~Pan.
\newblock Chamber structure and wallcrossing in the {ADHM} theory of curves
  {II}.
\newblock {\em J. Geom. Phys.}, 62(2):548--561, 2012.

\bibitem{flatGbundles}
K.~Corlette.
\newblock Flat {$G$}-bundles with canonical metrics.
\newblock {\em J. Differential Geom.}, 28(3):361--382, 1988.

\bibitem{hodge-char}
M.~de~Cataldo, T.~Hausel, and L.~Migliorini.
\newblock {Topology of Hitchin systems and Hodge theory of character
  varieties}.
\newblock arXiv:1004.1420, to appear in Ann. Math.

\bibitem{decompth}
M.~A.~A. de~Cataldo and L.~Migliorini.
\newblock The decomposition theorem, perverse sheaves and the topology of
  algebraic maps.
\newblock {\em Bull. Amer. Math. Soc. (N.S.)}, 46(4):535--633, 2009.

\bibitem{RelLefschetz}
M.~A.~A. de~Cataldo and L.~Migliorini.
\newblock The perverse filtration and the {L}efschetz hyperplane theorem.
\newblock {\em Ann. of Math. (2)}, 171(3):2089--2113, 2010.

\bibitem{Denef:2007vg}
F.~Denef and G.~W. Moore.
\newblock {Split states, entropy enigmas, holes and halos}.
\newblock {\em JHEP}, 1111:129, 2011.

\bibitem{modADHM}
D.~E. Diaconescu.
\newblock Moduli of {A}{D}{H}{M} sheaves and local {D}onaldson-{T}homas theory.
\newblock {\em J. Geom. Phys.}, (62):763--799.

\bibitem{Dijkgraaf:2006um}
R.~Dijkgraaf, C.~Vafa, and E.~Verlinde.
\newblock {M-theory and a topological string duality}.
\newblock 2006.
\newblock hep-th/0602087.

\bibitem{DG}
T.~Dimofte and S.~Gukov.
\newblock {Refined, Motivic, and Quantum}.
\newblock {\em Lett. Math. Phys.}, 91:1, 2010.

\bibitem{twistedharmonic}
S.~K. Donaldson.
\newblock Twisted harmonic maps and the self-duality equations.
\newblock {\em Proc. London Math. Soc. (3)}, 55(1):127--131, 1987.

\bibitem{EK-I}
T.~Eguchi and H.~Kanno.
\newblock {Five-dimensional gauge theories and local mirror symmetry}.
\newblock {\em Nucl. Phys.}, B586:331--345, 2000.

\bibitem{EK-II}
T.~Eguchi and H.~Kanno.
\newblock {Topological strings and Nekrasov's formulas}.
\newblock {\em JHEP}, 12:006, 2003.

\bibitem{GV-II}
R.~Gopakumar and C.~Vafa.
\newblock {{M} theory and topological strings {I}{I}}.
\newblock arXiv:9812127.

\bibitem{HRV}
T.~Hausel and F.~Rodriguez-Villegas.
\newblock Mixed {H}odge polynomials of character varieties.
\newblock {\em Invent. Math.}, 174(3):555--624, 2008.
\newblock With an appendix by Nicholas M. Katz.

\bibitem{hitchin-selfd}
N.~J. Hitchin.
\newblock The self-duality equations on a {R}iemann surface.
\newblock {\em Proc. London Math. Soc. (3)}, 55(1):59--126, 1987.

\bibitem{Hollowood:2003cv}
T.~J. Hollowood, A.~Iqbal, and C.~Vafa.
\newblock {Matrix Models, Geometric Engineering and Elliptic Genera}.
\newblock {\em JHEP}, 03:069, 2008.

\bibitem{HST}
S.~Hosono, M.-H. Saito, and A.~Takahashi.
\newblock Relative {L}efschetz action and {BPS} state counting.
\newblock {\em Internat. Math. Res. Notices}, (15):783--816, 2001.

\bibitem{algCS}
Z.~Hua.
\newblock {Chern-Simons functions on toric Calabi-Yau threefolds and
  Donaldson-Thomas theory}.
\newblock arXiv:1103.1921.

\bibitem{huylehn}
D.~Huybrechts and M.~Lehn.
\newblock {\em The geometry of moduli spaces of sheaves}.
\newblock Aspects of Mathematics, E31. Friedr. Vieweg \& Sohn, Braunschweig,
  1997.

\bibitem{fivedtwo}
K.~A. Intriligator, D.~R. Morrison, and N.~Seiberg.
\newblock {Five-dimensional supersymmetric gauge theories and degenerations of
  Calabi-Yau spaces}.
\newblock {\em Nucl.Phys.}, B497:56--100, 1997.

\bibitem{IKP-I}
A.~Iqbal and A.-K. Kashani-Poor.
\newblock {Instanton counting and Chern-Simons theory}.
\newblock {\em Adv. Theor. Math. Phys.}, 7:457--497, 2004.

\bibitem{IKP-II}
A.~Iqbal and A.-K. Kashani-Poor.
\newblock {SU(N) geometries and topological string amplitudes}.
\newblock {\em Adv. Theor. Math. Phys.}, 10:1--32, 2006.

\bibitem{Iqbal:2007ii}
A.~Iqbal, C.~Kozcaz, and C.~Vafa.
\newblock {The refined topological vertex}.
\newblock {\em JHEP}, 10:069, 2009.

\bibitem{IKV}
A.~Iqbal, C.~Kozcaz, and C.~Vafa.
\newblock {The refined topological vertex}.
\newblock {\em JHEP}, 10:069, 2009.

\bibitem{KMV}
S.~Katz, P.~Mayr, and C.~Vafa.
\newblock {Mirror symmetry and exact solution of 4-D N=2 gauge theories: 1.}
\newblock {\em Adv.Theor.Math.Phys.}, 1:53--114, 1998.

\bibitem{spinBH}
S.~H. Katz, A.~Klemm, and C.~Vafa.
\newblock {M-theory, topological strings and spinning black holes}.
\newblock {\em Adv. Theor. Math. Phys.}, 3:1445--1537, 1999.

\bibitem{geom_eng}
S.~H. Katz and C.~Vafa.
\newblock {Geometric engineering of N=1 quantum field theories}.
\newblock {\em Nucl.Phys.}, B497:196--204, 1997.

\bibitem{Konishi-I}
Y.~Konishi.
\newblock {Topological strings, instantons and asymptotic forms of
  Gopakumar-Vafa invariants}.
\newblock hep-th/0312090.

\bibitem{wallcrossing}
M.~Kontsevich and Y.~Soibelman.
\newblock Stability structures, {D}onaldson-{T}homas invariants and cluster
  transformations.
\newblock arXiv.org:0811.2435.

\bibitem{Lawrence:1997jr}
A.~E. Lawrence and N.~Nekrasov.
\newblock Instanton sums and five-dimensional gauge theories.
\newblock {\em Nucl. Phys.}, B513:239--265, 1998.

\bibitem{LLZ}
J.~Li, K.~Liu, and J.~Zhou.
\newblock Topological string partition functions as equivariant indices.
\newblock {\em Asian J. Math.}, 10(1):81--114, 2006.

\bibitem{MNOP-I}
D.~Maulik, N.~Nekrasov, A.~Okounkov, and R.~Pandharipande.
\newblock Gromov-{W}itten theory and {D}onaldson-{T}homas theory. {I}.
\newblock {\em Compos. Math.}, 142(5):1263--1285, 2006.

\bibitem{MPT}
D.~Maulik, R.~Pandharipande, and R.~Thomas.
\newblock {Curves on K3 surfaces and modular forms}.
\newblock {\em J. Topology}, (3):937--996, 2010.

\bibitem{MY}
D.~Maulik and Z.~Yun.
\newblock Macdonald formula for curves with planar singularities.
\newblock arXiv:1107.2175.

\bibitem{MS}
L.~Migliorini and V.~Shende.
\newblock A support theorem for {H}ilbert schemes of planar curves.
\newblock arXiv:1107.2355.

\bibitem{fivedone}
D.~R. Morrison and N.~Seiberg.
\newblock {Extremal transitions and five-dimensional supersymmetric field
  theories}.
\newblock {\em Nucl.Phys.}, B483:229--247, 1997.

\bibitem{adhm-rec}
S.~Mozgovoy.
\newblock {Solution of the motivic ADHM recursion formula}.
\newblock arXiv:1104.5698.

\bibitem{Nekrasov:2002qd}
N.~A. Nekrasov.
\newblock {Seiberg-Witten Prepotential From Instanton Counting}.
\newblock {\em Adv. Theor. Math. Phys.}, 7:831--864, 2004.

\bibitem{OP}
A.~Okounkov and R.~Pandharipande.
\newblock The local {D}onaldson-{T}homas theory of curves.
\newblock {\em Geom. Topol.}, (14):1503--1567, 2010.

\bibitem{stabpairs-I}
R.~Pandharipande and R.~P. Thomas.
\newblock Curve counting via stable pairs in the derived category.
\newblock {\em Invent. Math.}, 178(2):407--447, 2009.

\bibitem{stabpairs-III}
R.~Pandharipande and R.~P. Thomas.
\newblock Stable pairs and {BPS} invariants.
\newblock {\em J. Amer. Math. Soc.}, 23(1):267--297, 2010.

\bibitem{Higgslocal}
C.~T. Simpson.
\newblock Higgs bundles and local systems.
\newblock {\em Inst. Hautes \'Etudes Sci. Publ. Math.}, (75):5--95, 1992.

\bibitem{Hilb_DT_PT}
J.~Stoppa and R.~P. Thomas.
\newblock Hilbert schemes and stable pairs: {GIT} and derived category wall
  crossings.
\newblock {\em Bull. Soc. Math. France}, 139(3):297--339, 2011.

\bibitem{Toda_DT_PT}
Y.~Toda.
\newblock Curve counting theories via stable objects {I}. {DT}/{PT}
  correspondence.
\newblock {\em J. Amer. Math. Soc.}, 23(4):1119--1157, 2010.

\bibitem{generating_wallcrossing}
Y.~Toda.
\newblock Generating functions of stable pair invariants via wall-crossings in
  derived categories.
\newblock In {\em New developments in algebraic geometry, integrable systems
  and mirror symmetry ({RIMS}, {K}yoto, 2008)}, volume~59 of {\em Adv. Stud.
  Pure Math.}, pages 389--434. Math. Soc. Japan, Tokyo, 2010.

\bibitem{detmap}
Y.~Yuan.
\newblock Determinant line bundles on moduli spaces of pure sheaves on rational
  surfaces and strange duality.
\newblock arXiv.org:1005.3201.

\end{thebibliography}
 \bibliographystyle{abbrv}
\end{document}